\documentclass[12pt,preprint]{aastex}
\def\Deg{${}^\circ$\llap{.}}
\def\Min{${}^{\prime}$\llap{.}}
\def\Sec{${}^{\prime\prime}$\llap{.}}

\slugcomment{To appear in The Astrophysical Journal Letters}

\shorttitle{The outer edge of the LMC}
\shortauthors{Gallart et al.}

\begin{document}


\title{Surface Brightness and Stellar Populations at the Outer Edge of 
the Large Magellanic Cloud: No Stellar Halo Yet}

\author{Carme Gallart\altaffilmark{1}}
\affil{Instituto de Astrof\'\i sica de Canarias. 38200 La
Laguna. Tenerife, Canary Islands. Spain.}
\email{carme@iac.es}

\author{Peter B. Stetson}
\affil{Dominion Astrophysical Observatory, Herzberg Institute of Astrophysics, National Research Council of Canada, 5071 West Saanich Road, Victoria,
BC~V9E~2E7, Canada}\email{Peter.Stetson@nrc-cnrc.gc.ca}

\author{Eduardo Hardy\altaffilmark{2,}\altaffilmark{3}}
\affil{National Radio Astronomy Observatory. Casilla 36-D, Santiago, Chile}
\email{ehardy@nrao.edu}

\author{Frederic Pont}
\affil{Observatoire de Gen\`eve, CH-1290 Sauverny, Switzerland}
\email{Frederic.Pont@obs.unige.ch}

\and

\author{Robert Zinn} \affil{Department of Astronomy, Yale University,
P.O. Box 208101, New Haven, CT 06520-8101} \email{zinn@astro.yale.edu}


\altaffiltext{1}{Ram\'on y Cajal Fellow.}
\altaffiltext{2}{The National Radio Astronomy Observatory is a facility of the National Science Foundation operated under cooperative agreement by Associated Universities, Inc.}
\altaffiltext{3}{Adjoint Professor, Astronomy Dept. Universidad de Chile}

\begin{abstract}
We present a high quality CMD for a 36\arcmin $\times$ 36\arcmin\
field located 8\degr\ ($\simeq 7$ kpc) from the LMC center, as well as
a precise determination of the LMC surface brightness derived from the
resolved stellar population out to this large galactocentric
radius. This deep CMD shows for the first time the detailed age
distribution at this position, where the surface brightness is $V
\simeq$ 26.5 mag/\sq$\arcsec$. At a radius R$\simeq$ 474\arcmin\ the
main sequence is well populated from the oldest turnoff at $I \simeq
21.5$ to the 2.5 Gyr turnoff at $I \simeq 19.5$. Beyond this radius, a
relatively strong gradient in the density of stars with ages in the
$\simeq$ 2.5--4 Gyr range is apparent.
%, with very few stars in this age range beyond $R = 480$\arcmin.
There are some stars brighter and bluer than the main
population, quite uniformly distributed over the whole area surveyed,
which are well matched by a 1.5 Gyr isochrone and may be indicative of
a relatively recent star formation, or merger, event.
% which extended to this very external LMC position.
The surface brightness profile of the LMC remains exponential
to this large galactocentric radius and shows no evidence of disk
truncation. Combining the information on surface brightness and
stellar population we conclude that the LMC disk extends (and dominates
over a possible stellar halo) out to a distance of at least 7 kpc.
% is unlikely to have an old stellar halo similar to that of the Milky Way, 
% unless it appeared at an even larger galactocentric radius. 
These results confirm that the absence of blue stars in the relatively
shallow off-center CMDs of dIrr galaxies is not necessarily evidence
for an exclusively old stellar population resembling the halo of the
Milky Way.

\end{abstract}




\section{Introduction} \label{intro}

The nature of the stellar populations at the outer edges of dwarf
irregular (dIrr) galaxies, and in particular whether these galaxies
have a truly old stellar halo has been a subject of some controversy
over the last few years. The presence of a background sheet of red
stars, extending beyond the most prominent blue, young stellar
population in IC1613 and other dIrr galaxies was discussed already by
Baade (Sandage 1971). But, what is the nature of this sheet of red
stars? 
%Hardy et al.\ (1984) have shown that the underlying red
%population of the {\it central} regions of the LMC is mainly composed
%of intermediate--age stars, without a significant contribution from a
%superimposed halo-like population. 
Minniti \& Zijlstra (1996) claimed that the Local Group dIrr WLM has
an old (as old as the Milky Way globular clusters) stellar halo, from
the sole evidence that the color-magnitude diagram (CMD) of an outer
field extending out to 8\arcmin\ from the center of the galaxy, along
its minor axis, does not contain blue stars down to I $\simeq$ 23
($M_I\simeq-1.8$), and shows a well defined tip of the RGB. However,
since all stars older than about 1 Gyr produce a RGB at very
approximately the same location, and main--sequence (MS) stars older
than a few hundred Myr would be below the limit of Minniti \&
Zijlstra's photometry, all that can be confidently concluded from the
lack of bright MS stars is that the stellar population is older than a
few hundred Myr (see also the discussion by Aparicio, Tikhonov \&
Karachentsev 2000). Other evidence, such as the presence of C stars in
the outer regions of dwarf galaxies (e.g. Letarte et al.\ 2002), or a
full modeling of the CMD using synthetic CMDs (Aparicio \& Tikhonov
2000) suggests in fact the presence of a range of ages.

Determining the age structure and kinematic properties of the outer
red component of dIrr galaxies may hold important clues on the process
of dwarf galaxy formation and evolution. Do these stars represent a
stellar component truly differentiated from the younger one, with the
characteristics of a halo population similar to that of the Milky Way
(old, metal poor, devoid of gas, kinematically hot), or is there a
smooth gradient of stellar ages, which might suggest that star
formation has been ``shrinking'' with time? (Hidalgo, Mar\'\i n-Franch
\& Aparicio 2003)

A complete answer to these questions requires both very deep
photometry, reaching the oldest MS turnoffs, and relatively
high dispersion spectroscopy to obtain kinematics of individual
stars. From deep CMDs at different galactocentric distances it is
possible to characterize the ages of the stellar populations present,
and to assess, therefore, whether a ``break'' in the stellar
population, indicating an edge to the zone of protracted star
formation and a transition to an old stellar halo actually occurs. The
observational requirements for this task are, however,
challenging. For the Milky Way satellites, it is necessary to reach $V
\simeq 24$ to obtain good precision photometry of the old MS turnoff
region. For the rest of the Local Group, $V\simeq 28$ is
necessary. While the first is attainable from the ground, the latter
is only possible using the HST equipped with ACS. Obtaining the
required spectroscopic information is a highly challenging
enterprise.

The Magellanic Clouds are the two nearest Irr galaxies. Their old MS
turnoffs are located near $V\simeq$ 23. While this magnitude is easily
attainable from the ground, the huge angular sizes of the Clouds makes
the mapping of their full extent and the characterization of their
stellar population gradients a daunting task. As discussed by Irwin
(1991) on the basis of isopleth maps, the LMC extends over an
elliptical area covering 23\degr\ by 17\degr, with the major axis
oriented roughly in the North-South direction. This is also basically
the extent of the LMC cluster system. On the North side, North of
declination --62\degr\ ($\simeq 8\degr$ from the center), the isopleths
abruptly separate from each other. Does this indicate some kind of
break in the characteristics of the stellar populations? Some
information about the ages of the stars present at this position comes
from the study of Kunkel, Irwin \& Demers (1997), who searched for C
stars in the periphery of the two Magellanic Clouds. Interestingly,
they found a relatively sharp boundary for the outermost distribution
of C stars at a radius of $\simeq 8\degr$ along the major axis, which
may indicate the actual maximum extent of an intermediate-age
population. A detailed characterization of such a population, however,
requires obtaining CMDs reaching the oldest MS turnoffs. A full
dynamical study of the C star population of the LMC based on 1043
objects out to $8 \degr$ from the center was published by van der
Marel et al.\ (2002), who inferred the presence of a dark-matter halo
containing roughly 2/3 of the total mass of the LMC. The presence of a
kinematically hot, metal-poor, old stellar halo has been inferred by
Minniti et al.\ (2003) from the kinematics of a sample RR Lyrae stars.

%The HI rotation curve has been measured out to a radius of 3.5 \degr only.

With the purpose of characterizing the stellar population gradients in
the LMC, we imaged four fields situated at distances ranging from 3.1
to 8 degrees from the LMC center using the CTIO 4m MOSAIC wide field
imager. The full analysis of the whole dataset is underway (Gallart et
al.\ 2004). In this Letter we analyze the CMD of the outermost field,
which shows a somewhat surprising result on the age composition of the
outer LMC, and use the whole dataset to trace the surface brightness
across this galaxy.

\section{Observations and data reduction.} \label{obs}

We obtained V and I images of a LMC field centered at 7\Deg8 from
its center to the North ($\alpha:$ 05:13:17; $\delta:$ --61:58:40) on
January 17th and 18th, 2001, using the Mosaic II CCD Imager on the
CTIO Blanco 4m telescope. The Mosaic II provides a total field of
36\arcmin $\times$ 36\arcmin, with a resolution of 0\Sec27/pixel.
Seeing was typically around 1\Sec0. Exposure times were
6$\times$900$\,$s in V {\it plus} 8$\times$600$\,$s in I. Several short
exposures were also obtained to retrieve the photometry of the
brightest stars. The nights were photometric, and the Landolt (1992)
standard star fields SA92, SA95, SA98, SA104 were observed several
times each night for calibration purposes. We used Stetson's (2000)
photometry for these fields to calibrate our photometry. During the
same run and in another one in December 1999, we obtained photometry
of similar quality for three more internal fields. These results will
be described elsewhere (Gallart et al.\ 2004), and used occasionally
here (Section \ref{location}).

The Mosaic frames were reduced in a standard way, as indicated in the
{\it CTIO Mosaic II Imager User Manual} (Schommer et al.\ 2000), using
the MSCRED package within IRAF.  Finally, profile-fitting photometry
of the LMC fields was obtained using the DAOPHOT / ALLFRAME suite of
computer programs (Stetson 1987, 1994). A full description of the data
reduction and photometry is provided in Gallart et al.\ (2004).

\section{The LMC surface brightness profile: reaching the outer
edge?}\label{location}

We have derived the surface brightness for the four fields we observed
in the LMC, by integrating the flux in the CMD\footnote{The
contribution to the total flux of the unresolved LMC population
(i.e. by the stars fainter than $M_V=5$) may range between 1 and 10\%
of the total flux, as estimated from synthetic CMDs for a composite
population (constant star formation rate from 15 Gyr ago to the
present time) and for a purely old population. We estimate, therefore,
that we may be underestimating the surface brightness by a maximum of
0.1 magnitudes.}. This method has the advantage that a very accurate
field subtraction can be performed at the very low surface-brightness
regime as foreground stars and background galaxies are for the most
part well separated from the characteristic sequences of the LMC
CMD. This is particularly important in the outermost field, where the
flux contribution of Milky Way foreground objects exceeds by far that
of the galaxy. The process will be fully described in Gallart et
al.\ (2004). In short, we used the predictions of the Besan\c{c}on Galaxy
model (Robin et al.\ 2003), but scaling the number of predicted Milky
Way stars to the number observed in our field in zones in the CMD
clearly devoid of LMC stars. Without this scaling, we found that
Besan\c{c}on models predicted 25\% more Galactic stars than we observed
at relatively bright magnitudes. In addition, we only considered stars
fainter than $V=15$ when integrating the flux in the field, since no
stars belonging to the LMC were observed brighter than this
limit. Since field stars brighter than $V=15$ are the main contributor
to the total flux of the field, this greatly reduced the error in its
subtraction. We neglected the contribution of background unresolved
galaxies, since they are faint.

\begin{figure}
\plotone{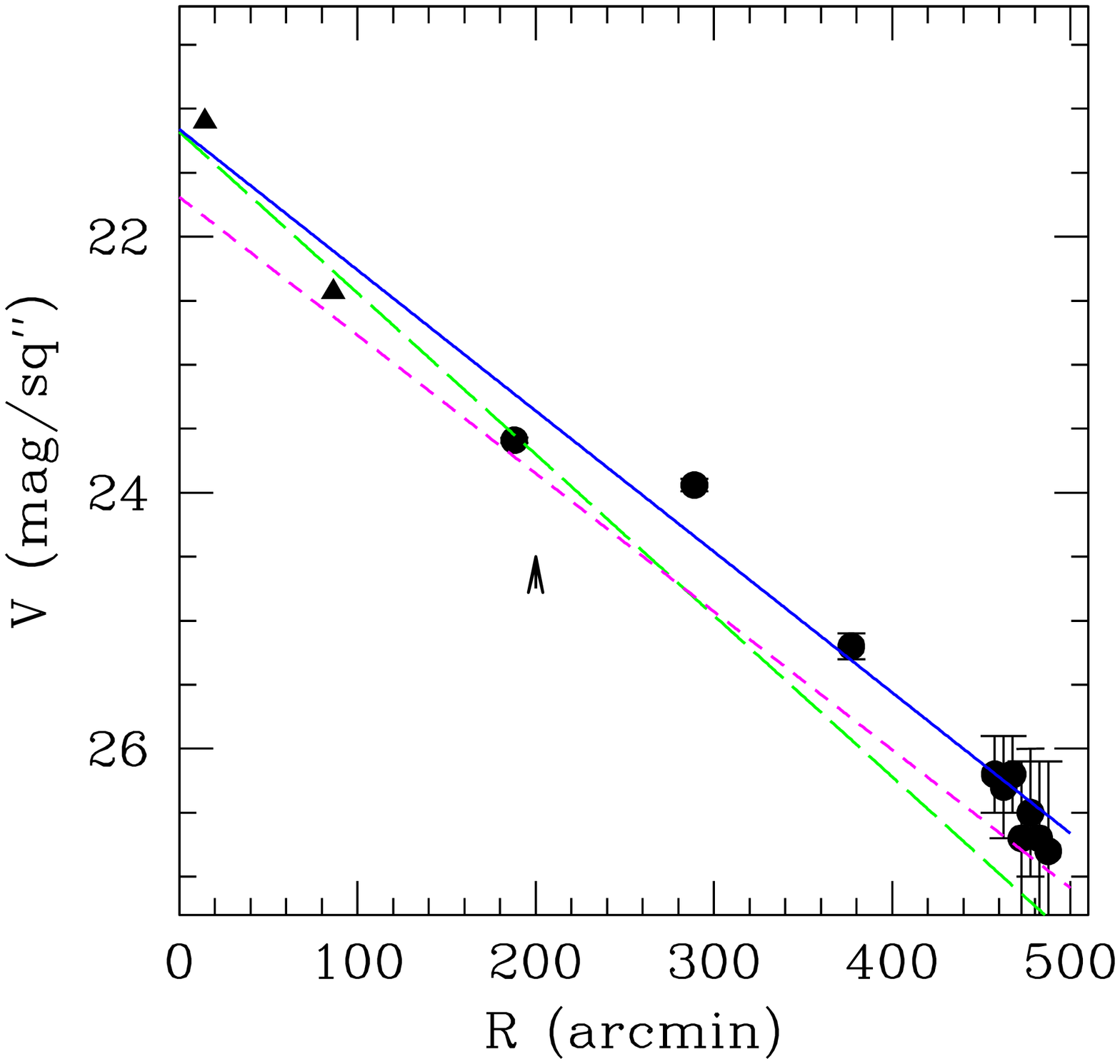}
\caption{$V$ (dots) surface brightness profile of the LMC as measured 
from our MOSAIC data. The field discussed in this paper is the
outermost one. The surface brightness has been calculated across this
field and given for seven sections at increasing distances from the center
of the LMC. Two measurements of Hardy (1978, his innermost point and
the outermost one of his ``outer regions'') have been added as
triangles.  The outermost radius measured by Bothun \& Thompson (1988)
is marked with an arrow, which also indicates an estimate of the $V$
surface brightness measured by them at this radius, assuming
$(B-V)=0.5$. Exponential laws for the LMC disk as calculated from the
whole dataset represented in the figure (Hardy 1978 {\it plus} our
data, solid line; $\mu_i(0)$=21.16, $\alpha$=98\Min7) and as
derived by Hardy (1978, long-dashed line; $\mu_i(0)$=21.18,
$\alpha$=86\Min2) and Bothun \& Thompson (1988, short-dashed line;
$\mu_i(0)$=21.16, $\alpha$=100\Min6) are represented.
\label{brillosup}}
\end{figure}

In Figure \ref{brillosup}, the surface brightness of our fields
measured in V (dots) is represented as a function of distance from the
center of mass of the LMC. We have divided the outermost field into seven
sections at increasing distances from the center of the LMC and
calculated the surface brightness for each, in order to check whether
the surface brightness variation remained smooth (which seems to be
the case) or whether some kind of truncation was observed instead. The
error bars were estimated by repeating the calculations with sky
backgrounds that varied by 25\% from our best estimate.  We have added
(triangles) two measurements from Hardy (1978). Exponential laws for the
LMC disk as calculated from the whole dataset (Hardy {\it plus} our
data, solid line), and as derived by Hardy (long-dashed line) and
Bothun \& Thompson (1988, short-dashed line) are represented. Our fit
has a slope similar to that given by Bothun \& Thompson, but is
brighter.  We also don't see evidence of the truncation found by these
authors at a galactocentric distance of approximately 170\arcmin. We
conclude that the surface-brightness measurements displayed are
consistent with an exponential law out to the last point measured.

The field we will discuss in this Letter is the outermost one,
situated from 450\arcmin\, to 490\arcmin\, from the center of the
LMC. The surface brightness in this field ranges from 26.2 to 26.8
mag/\sq\arcsec in $V$ and is one magnitude brighter in $I$. To put
the location of our field in context, we have drawn an arrow at the
location of the outermost radius for which the LMC surface brightness
was measured by Bothun \& Thompson (1988). The tip of the arrow
indicates an estimate of the $V$ magnitude measured by these authors
at this position, calculated using their $B$ magnitude and assuming a
color $(B-V)=0.5$. This location also approximately coincides with the
outer radius of the HI measured by Kim et al.\ (1998). The radius of the
outermost field that we have measured is roughly equal to the
radii of the maximum extension of the LMC cluster system (Olszewski et
al.\ 1988) and the outer edge of the C-star system obtained by Kunkel
et al.\ (1997).

\section{The Color-Magnitude Diagrams} \label {cmds}

\begin{figure}
\plotone{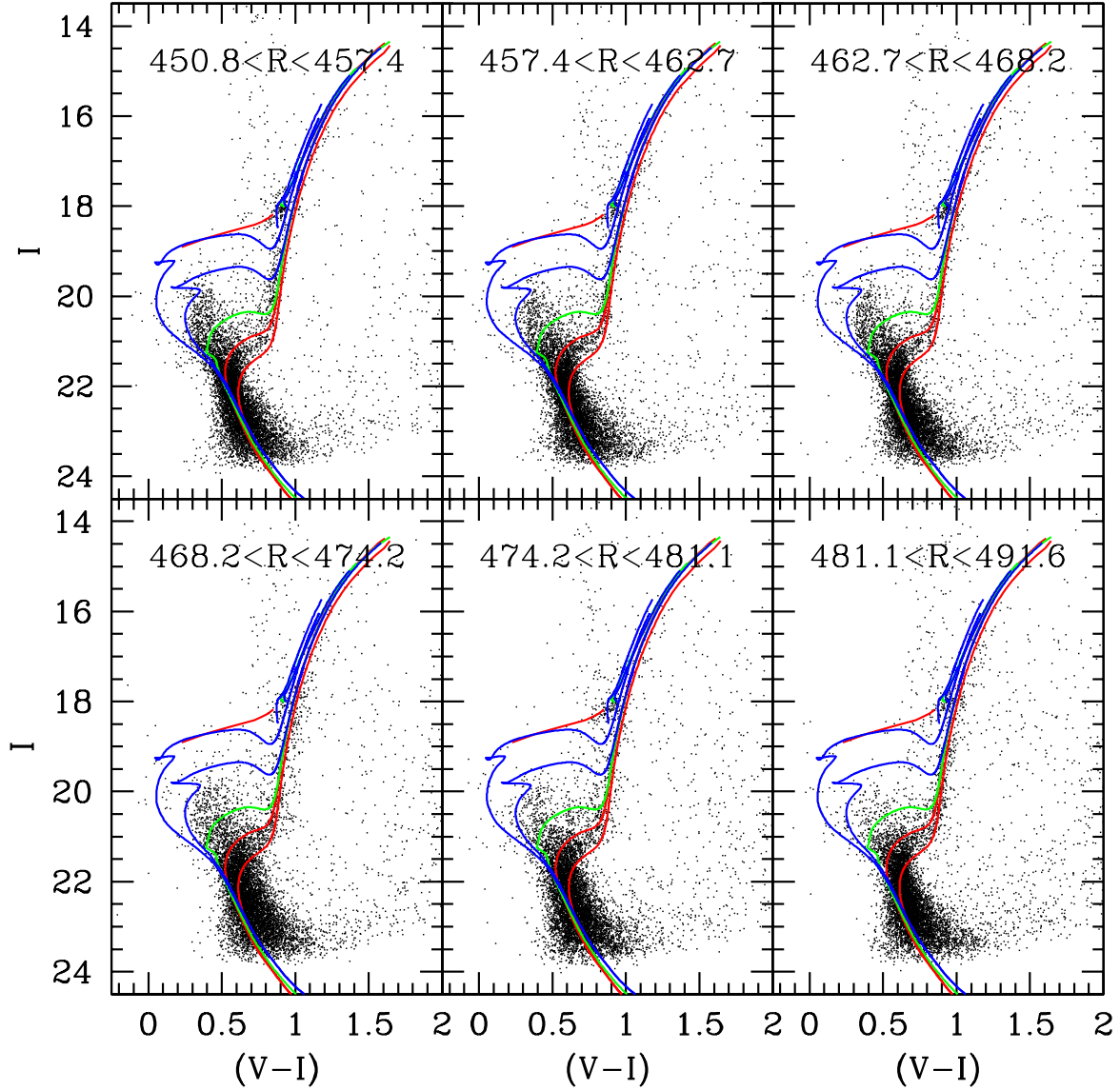}
\caption{Observed [(V-I), I] CMDs in order of increasing distance from the
LMC center. The number of stars in each CMD is the same and, since the
density of LMC stars decreases as a function of radius, $\delta R$
varies for each CMD. The range in R of each CMD is labeled. Isochrones
by Pietrinferni et al.\ (2004) for Z=0.001 and ages 13.5 and 8 Gyr
(red), Z=0.002 and age 4.5 Gyr (green) and Z=0.004 and ages 2.5 and
1.5 Gyr (blue) have been superimposed.\label{dcmad_iso}}
\end{figure}

Figure~\ref{dcmad_iso} shows $[(V-I),I]$ CMDs of the LMC, in order of
increasing distance, R, from the LMC center, from galactocentric radius
450 to 490\arcmin\ (see caption for details). Stars with at least one
valid measurement in each filter have been selected using the
following limits for the error and shape parameters given by ALLFRAME:
$\sigma_{(V-I)} (= \sqrt{\sigma_V^2+\sigma_I^2}) \le 0.1$, $\mid sharp
\mid \le 0.3$ and $\chi \le 4.5$.  A total of 55000 stars down to
$I\leq24.0$ have been measured. The CMDs reach a couple of magnitudes
below the old MS turnoff and therefore, this point is
always cleanly reached with good photometric accuracy. Isochrones from
Pietrinferni et al.\ (2004) for Z=0.001 and ages 13.5 and 8 Gyr,
Z=0.002 and age 4.5 Gyr and Z=0.004 and ages 2.5 and 1.5 Gyr have been
superimposed.

These CMDs show for the first time the details of the age structure of
the stellar population at the outer edge of the LMC. Out to a radius
R$\simeq$474\arcmin\, the MS appears populated quite smoothly from the
oldest turnoff at $I \simeq 21.5$ to the 2.5 Gyr turnoff at $I \simeq
19.5$. Beyond this radius, a relatively strong gradient in the
density of stars of ages $\simeq$ 4-2.5 Gyr begins to be apparent,
with very few stars in this range beyond R=480\arcmin. There is a
number of stars brighter and bluer than the main population, which are
well matched by a 1.5 Gyr isochrone. They are distributed quite
uniformly over the whole field. No foreground stars or background
galaxies, however, are expected to populate the CMD at this position,
and therefore they must be young stars genuinely belonging to the
LMC. They may indicate a small burst of star formation 1.5 Gyr ago,
the spatial extent of which cannot be evaluated with the current
dataset. A survey of a larger area, and more fields at similar radius
but different position angles would be necessary to ascertain whether
it has been a global or a localized event. The fainter stars in the
subgiant-branch region are well matched by the 13.5 Gyr, Z=0.001
isochrone, but no blue extended horizontal branch can be seen in the
CMDs. The red clump---or red horizontal branch---presents, however, a
short faint blue extension.

\begin{figure}
\plotone{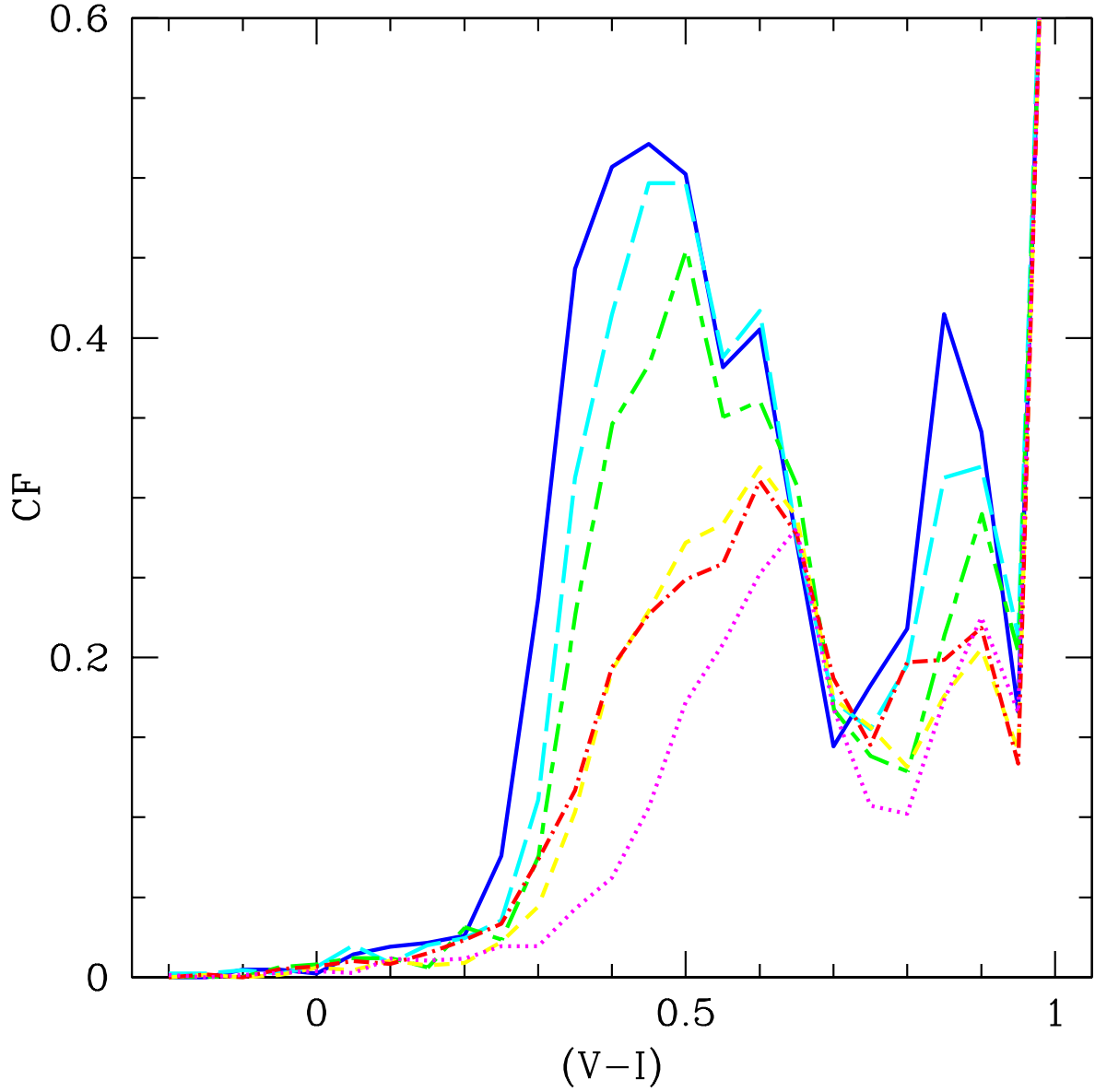}
\caption{Color function of the CMD
integrated over the magnitude range $21.5\le I \le 17.5$, for each of
the CMDs shown in Figure \ref{dcmad_iso}. Solid,
long-dashed, short-long dashed, short-dashed, dot-short dashed and
dotted lines are in order of increasing galactocentric distance,
respectively (see text for details). \label{cfms6_norma}}
\end{figure}

Figure \ref{cfms6_norma} displays the color function of the CMD
integrated over the magnitude range $21.5\le I \le 17.5$,
i.e. including all stars from just below the red clump to the location
of the oldest MS turnoffs.  The star counts have been normalized to
the number of stars in the last two bins, which mainly include the
red clump and a portion of the RGB just below it. Since these
structures contain stars of ages greater than $\simeq$ 1 Gyr, this is
equivalent to a normalization by an indicator of the mean star
formation rate. The gradient in the number of blue stars present in
these CMDs, and their color distribution is very striking, notably
more pronounced than would be expected from a visual inspection of the
CMDs alone. The main population of age $\ge 2.5$ Gyr present in the
CMD appears at $(V-I)
\simeq 0.2$, and the non-negligible number of stars between
$(V-I)=0-0.2$ indicate the presence of the population of age $\simeq
1.5$ Gyr mentioned above (error-bars have been omitted for clarity,
but they are smaller than the differences between the curves). Stars
bluer than $(V-I) \simeq 0.5$ are younger than $\simeq$ 8 Gyr, and it
is in this interval of color that the largest differences between
fields are apparent. Beyond R$\simeq$470\arcmin\ the population of
stars younger than 8 Gyr starts to abruptly decrease. The interval
$(V-I)=0.6-0.7$ is dominated by stars older than 8 Gyr, for which the relative
numbers are very similar at all radii. Differences in the color
function appear again in the interval $(V-I)=0.8-0.9$, which mainly
contains the subgiant branches of populations of all ages.

\section{Discussion}

We have presented a high-quality CMD, reaching the oldest main
sequence turnoffs, of an LMC field situated at R=450--490\arcmin\ from
the center of mass of the LMC, toward the North (i.e., in the
direction of its major axis). Very few CMDs of the LMC field have been
published beyond this radius. The most targeted, detailed work is, to
our knowledge, that of Stryker (1984) who did photographic photometry
of stars in an area of 118\Min4 $\times$ 118\Min4 situated
9\degr\ northeast of the LMC bar. She just reached the oldest
MS turnoffs. In spite of the very large photometric errors,
she was able to show that a blue population brighter than the old
turnoff was indeed present. In the very large area surveyed, she was
also unable to find evidence of a blue horizontal branch.

In this paper, we show the presence of a substantial intermediate-age
stellar population at R = 8\degr \,($\sim$ 7 kpc) from the center of
the LMC. A strong gradient in this population, however, occurs in a
range of just 0.5 kpc. The combination of an exponential surface
brightness profile together with intermediate-age stars indicates the
presence of a dominant disk population even at these large
galactocentric distances.  It would be of great interest to
investigate the stellar populations at even larger radial distances,
to ascertain whether eventually a background old population appears
and stays for a while (possibly indicating a true old {\it stellar}
halo population that would need to be investigated further), or
whether the actual outer edge of the LMC is smoothly reached with a
continuing gradient of stellar populations. The lack of a well
populated, blue extended horizontal branch, however, both in the field
investigated here and in the larger, even more peripheral field
studied by Stryker (1984) seems to indicate that the amount of {\it
field} stellar population as old and metal poor as that of the Milky
Way halo globular clusters and dSph galaxies is small in the LMC.

%goes against the presence of a stellar
%halo even at these large galactocentric distances. At these distances
%the halo/disk contrast should be significant. The disk--like
%kinematics of the intermediate--age population as typified by the C
%stars studied by van der Marel et al.\ (2002) also argues against a
%stellar halo at this distance. We conclude that is is unlikely that
%the LMC posseses a stellar halo, although the evidence for a dark one
%is strong as shown by van der Marel et al.\ (2002).

Turning now to the origin of the intermediate-age population in the
outer fields, it is important to ask whether it could have formed at
its current position, or it must have migrated from a more internal
birth site (Hodge, Hunter \& Oey 2003). Since the rotation of the
galaxy may make it difficult for stars to move outwards, this
population of stars may be a sign that the HI gas extended as far as
this outer field as recently as a few Gyr ago and has since
(gradually, thus the observed gradient), receded to the more central
regions. Regarding the youngest $\simeq$ 1.5 Gyr old population, which
seems to be quite uniformly distributed over the field, an interaction
between the LMC and the SMC, or the merger of the LMC with a smaller
galaxy are attractive mechanisms for either in-situ formation, or
accretion of stars in an apparently discrete age range.

Our results, in any case, are a demonstration that the lack of MS
stars and the presence of a RGB with characteristics very similar to
that of an old population cannot be taken as a proof of the existence
of an exclusively old stellar population. The presence of a true
stellar halo in dIrr galaxies remains therefore unproven, and it needs
to be investigated with much deeper CMDs aided with spectroscopic data
to characterize robustly the metallicities and the kinematics of the
individual stars.

\acknowledgments

This research is part of a joint project between the Universidad de
Chile and Yale University that is partially funded by the Fundaci\'on
Andes.  C.G. acknowledges partial support from the IAC (Project
3I1902), the Spanish Ministry of Science and Technology (Plan Nacional
de Investigaci\'on Cient\'\i fica, Desarrollo e Investigaci\'on
Tecnol\'ogica, AYA2002-01939) and from the European Structural
Funds. R.Z. acknowledges the support of NSF grant AST-0098428.

%% In this section, we use  the \subsection command to set off
%% a subsection.  \footnote is used to insert a footnote to the text.

%% The \notetoeditor{TEXT} command allows the author to communicate
%% information to the copy editor.  This information will appear as a
%% footnote on the printed copy for the manuscript style file.  Nothing will
%% appear on the printed copy if the preprint or
%% preprint2 style files are used.


%% Included in this acknowledgments section are examples of the
%% AASTeX hypertext markup commands. Use \url without the optional [HREF]
%% argument when you want to print the url directly in the text. Otherwise,
%% use either \url or \anchor, with the HREF as the first argument and the
%% text to be printed in the second.

%\appendix

\clearpage


%% If you are not including electonic art with your submission, you may
%% mark up your captions using the \figcaption command. See the
%% No more than seven \figcaption commands are allowed per page,
%% so if you have more than seven captions, insert a \clearpage

%% Tables should be submitted one per page, so put a \clearpage before

%% Two options are available to the author for producing tables:  the
%% deluxetable environment provided by the AASTeX package or the LaTeX
%% table environment.  Use of deluxetable is preferred.
%%

%% Three table samples follow, two marked up in the deluxetable environment,
%% one marked up as a LaTeX table.

%% In this first example, note that the \tabletypesize{}
%% command has been used to reduce the font size of the table.
%% Note also that the \label command needs to be placed

%\clearpage

%% The following command ends your manuscript. LaTeX will ignore any text

\end{document}